# A Novel Scheduling Framework Leveraging Hardware Cache Partitioning for Cache-Side-Channel Elimination in Clouds

Read Sprabery
University of Illinois,
Urbana-Champaign
spraber2@illinois.edu

Konstantin Evchenko
University of Illinois,
Urbana-Champaign
evchenk2@illinois.edu

Abhilash Raj
Oregon State University
rajab@oregonstate.edu

Rakesh B. Bobba
Oregon State University
rakesh.bobba@oregonstate.edu

Sibin Mohan
University of Illinois,
Urbana-Champaign
sibin@illinois.edu

Roy H. Campbell
University of Illinois,
Urbana-Champaign
rhc@illinois.edu

## ABSTRACT

While there exist many isolation mechanisms that are available to cloud service providers, including virtual machines, containers, *etc.*, the problem of side-channel increases in importance as a remaining security vulnerability – particularly in the presence of shared caches and multicore processors. In this paper we present a hardware-software mechanism that improves the isolation of cloud processes in the presence of shared caches on multicore chips. Combining the Intel CAT architecture that enables cache partitioning on the fly with novel scheduling techniques and state cleansing mechanisms, we enable cache-side-channel free computing for Linux-based containers and virtual machines, in particular, those managed by KVM. We do a preliminary evaluation of our system using a CPU bound workload. Our system allows Simultaneous Multithreading (SMT) to remain enabled and does not require application level changes.

## 1 INTRODUCTION

Cache-based side-channel attacks (*e.g.,* [13, 26, 27]) are a threat to computing environments where a diverse set of users share hardware resources. Such attacks take advantage of observable side-effects (on hardware) due to the execution of software programs. A number of these attacks focus on differences in timing while accessing shared processor caches. Recently, researchers have adapted these cache-based side-channel attacks to cloud computing environments, especially Infrastructure-as-a-Service (IaaS) clouds (*e.g.,* [18, 23, 30, 47]), and showed that secrets and sensitive information can be extracted across co-located virtual machines (VMs). Container frameworks such as Docker[7] that virtualize the operating system are even more susceptible to such attacks since they share the underlying operating system (*e.g.,* [48]).

Initial cache-based side-channel attacks focused on schedulers, either at the OS level or Virtual Machine Monitor (VMM) layer [13, 26, 47]. Other approaches also focused on resource sharing – *e.g.,* using the processor core (for instance SMT on a single core) to access shared L1 and L2 caches [27]. Recent attacks have focused on the Last-Level-Cache (LLC) that is shared across multiple cores [18, 23, 43, 44, 48] – these make defenses much harder.

Many defenses against cache-side-channel attacks in cloud environments have also been proposed (*e.g.,* [6, 19, 21, 22, 24, 25, 28, 31, 32, 37–40, 45, 51]). However, the proposed solutions suffer from a variety of drawbacks: *(a)* some are probabilistic [25, 37, 51]; *(b)* others do not protect applications when SMT is enabled [51]; *(c)* some require developers to re-write applications [19, 22, 31], *(d)* while others require hardware changes [39, 40] impacting deployability; *(e)* some depend on violating x86 semantics by modifying the resolution, accuracy or availability of timing instructions [21, 24, 38] and consequently require significant changes to the applications. Compiler-based [6] and page coloring based cache-partitioning [28, 32, 45] approaches have high overheads making them impractical.

Defenses against cache-side-channel attacks that *eliminate* the attacks, rather than frustrate techniques employed by the attacker are desirable. Shannon's noisy-channel coding theorem states that information can be transmitted, regardless of the amount of noise on the channel. Probabilistic defenses (*e.g.,* [25, 37, 51]) may decrease the bit-rate of attacks, but cannot fully eliminate them. In addition to a guaranteed defense, the solution must not severely impact *(i)* the performance of the applications or *(ii)* utilization of the machine. In other words, defenses must minimize the performance impact of enforcing hard isolation to remain practical. For instance, disabling hyper-threading [51], which many existing solutions do, can have a significant performance impact on the applications. To the best of our knowledge, every cloud provider enables hyperthreading. Furthermore, the solutions must be *easy to adopt*. History has shown that solutions requiring additional development time (or significant changes to existing applications) have a harder time being adopted (as shown in the Return Oriented Programming (ROP) community [34]). Thus, solutions that require developers make application level changes [19, 22] may be challenging to apply to existing workloads.

In this paper, we present a framework designed to eliminate side-channel attacks in cloud computing systems that use multicore processors with a shared LLC. The proposed framework uses a combination of *Commodity-off-the-Shelf (COTS) hardware features* along with *novel scheduling techniques* to eliminate cache-based side-channel attacks. In particular, we use Cache Allocation Technology (CAT) [17] that allows us to *partition last-level caches at runtime*. CAT, coupled with *state cleansing* between context switches, and selective sharing of common libraries removes any possibility of cache-timing-based side-channel attacks between different security domains. We implement a novel scheduling method, as an extension to the commonly-used Completely-Fair-Scheduler (CFS) in Linux, to reduce the overheads inherent due to any such cleansing operation. Our solution provides a *transparent*[1] way to eliminate side-channel attacks, while still working with *hyperthreading enabled* systems. It works with containers, kernel-based virtual machines (KVMs) and any other schedulable entity that relies on the OS scheduler[2].

In summary, the proposed framework:

**C1** Eliminates cache-based side-channel attacks for schedulable units (*e.g.,* containers, KVMs)

**C2** Requires no application level changes

**C3** Allows providers to exploit hyperthreading, and

**C4** Imposes modest performance overheads

## 2 SYSTEM AND ATTACK MODEL

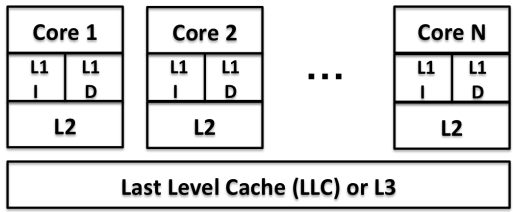

**Figure 1: Hierarchical Cache in Modern Processors**

### 2.1 System Model

We consider public Platform-as-a-Service (PaaS) or Infrastructure-as-a-cloud (IaaS) cloud environments. Such environments allow for co-residency of multiple computational appliances (*e.g.,* containers, VMs) belonging to potentially different security domains. We assume that the cloud computing infrastructure is built using commodity-off-the-shelf (COTS) components. In particular, we assume that the servers have multi-core processors with multiple levels of caches, some of which are shared (See Figure 1). We also assume that the servers have a runtime mechanism to partition the last-level shared cache.

For this work, we evaluated our approach using an Intel Haswell series processor that has a three-level cache hierarchy: private level 1 (L1) and level 2 (L2) caches for each core (64KB and 256KB respectively) and a last level (L3) cache (20MB) that is shared among all the cores[3]. For cache partitioning, we turned to the Intel *Cache Allocation Technology* (CAT)[17] that allows us to partition the shared L3 cache. The CAT mechanism is configured using model-specific registers (MSRs). This can be carried out at runtime in a dynamic fashion using software mechanisms. On our processor the maximum number of partitions is limited to *four* but newer generations support more [16]. Intel CAT technology has been available on select Haswell series processors starting late 2014 and continues to be available on select processor lines belonging to

Broadwell and Skylake micro-architectures that succeeded Haswell micro-architecture.

While our implementation and evaluation used an Intel Haswell processor with Intel CAT technology the proposed approach is generally applicable to any multi-core processor with a hierarchical cache, shared last-level cache, and a mechanism to partition the shared last-level cache.

### 2.2 Attack Model

While there are many potential threats to security in public cloud environments (*e.g.,* [5, 35]) the focus of this paper is on cache-based side-channel attacks (*e.g.,* [46, 48]). At a high-level, in cache-based side-channel attacks an attacker deduces information about the victim's computations by observing the victims cache usage. The information deduced can range from high-level information such as which tenant you are co-located with (*e.g.,* [30]) to more fine grained details such as cryptographic keys or items in a shopping cart (*e.g.,* [18, 48]).

In such attacks an attacker process first needs to co-locate (*i.e.,* get assigned to the same physical server) itself with the target or victim in the infrastructure. Methods to both achieve co-residency [30] and to thwart co-residency (*e.g.,* [1, 14, 15, 25, 49]) have been discussed in the literature. In this work we assume that an attacker is able to co-locate with the victim and focus on thwarting side-channel attacks themselves. Our framework complements approaches that thwart co-residency.

There are primarily two techniques to exploit cache based side-channels discussed in the literature, namely, 'Prime+Probe' attacks [27] and 'Flush-Reload" attacks [13]. It is important to note that while these techniques are popular, they are only possible because of measurable interference. Our solution addresses these specific techniques and other techniques that leverage the cache as a side-channel.

*(a)* **"Prime+Probe" attacks** [27]: At a high level, the attack 'primes' the entire cache with random data. After waiting for a certain time (so that the victim executes), the adversary then 'probes' the cache to see which of its own lines have been evicted – thus providing information on which lines were used by the victim. This information combined with the knowledge of cache access patterns exhibited by victim programs (*e.g.,* cryptographic algorithms) can be used to extract information (*e.g.,* cryptographic key being used) about the victim.

*(b)* **"Flush-Reload" attacks** [13]: are a modification of prime-and-probe attacks and leverage memory pages that are shared between the attacker and the victim. This allows the attacker to 'flush' *specific lines* rather than prime the whole cache. The rest of the procedure is similar – waiting for a certain amount of time to see which of the lines corresponding to shared memory pages are in the cache (hence used by the victim). This method is more efficient and results in less noise for the attacker.

We consider consider both cross-core (*i.e.,* attacker and victim running on different cores on the same processor) and same-core (*i.e.,* attacker and victim running on the same core) side-channel attacks. Same-core attacks, as the name indicates, require the attacker to achieve co-residency on the same core and be able to preempt the victim. They typically focus on higher-level caches (L1 and L2) that are specific to the core. Cross-core attacks on the other hand only require the attacker to achieve co-residency on the same physical server. Their limitation is that they only allow the attacker to observe victim's activities through the last level-cache which is shared and thus is noisy. However, it has been shown that such limitations can be overcome [23].

---

[1]From the perspective of the application developer/use

[2]For ease of exposition, in the rest of the paper, we will describe our framework using containers.

[3]This is the model of the system that we use for the rest of this paper.

To clarify, the attacker is capable of achieving co-residency with a victim, can allocate an arbitrary number of resources, and game both the cloud level scheduler (placement), and the operating system level scheduler (preemption). However, we assume that the cloud infrastructure is trusted. That is, while the cloud scheduler may be gamed, we assume that both the attacker and victim are authenticated with the cloud provider (e.g., for billing purposes). Additionally, we assume that the host kernel running either KVM or containers is trusted.

It is important to note that there are many other threats to cloud computing environments apart from cache-based side-channels. We limit the scope of this work to addressing cache-based side channels attacks. Such attacks can be launched by any legitimate tenant without having to exploit any vulnerabilities and without compromising the underlying software or hardware.

## 3 SYSTEM ARCHITECTURE

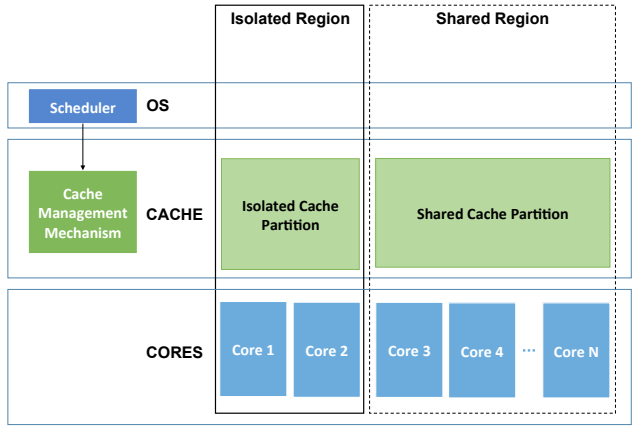

**Figure 2: System Overview**

Our framework logically partitions a host server into an *isolated region*[4] and a *shared region* as illustrated in Figure 2. Tenants are required to indicate to the cloud provider whether or not their containers need isolated execution. Containers designated as requiring isolated execution will be executed in the isolated partition of the host server, while all other containers will be executed in the shared partition. The 'isolated execution' designation guarantees that processes within the designated containers will not share cache resources with (i) processes from any container belonging to another tenant (or security domain), or with (ii) processes belonging to any container that is not designated for isolated execution irrespective of their ownership (or security domain).

Our design, discussed next, leverages (i) Intel CAT, processor affinities (or CPU pinning) and selective page sharing to provide spatial isolation, and (ii) co-scheduling with state cleansing to provide temporal isolation for designated containers.

### 3.1 Hardware Enforced Spatial Isolation

Intel's CAT [17], currently available in COTS hardware in Intel's Xeon series processors, is designed to improve the performance of latency sensitive real-time workloads by allowing the LLC to be partitioned into distinct regions. Each processor core can be restricted to allocating cache lines into a specified cache partition. Consequently, a processor can only evict cache lines within the processor's assigned LLC partition, thus reducing the impact of processes running on that processor core can have on other cache regions and vice versa. In particular, note that the ability to allocate cache lines (priming in prime+probe attacks) in a cache shared with the victim, and the ability to evict cache lines being used by the victim process (flushing in flush-reload attacks) are key steps in cache-based side channel attacks. Therefore ensuring that a potential victim and attacker processes run on cores associated with different LLC cache partitions, as we propose in our framework, defeats some cache-side-channel attacks. Specifically, cross-core prime+probe attacks on a victim process using a different cache partition are eliminated.

Accordingly, we partition the LLC into two regions using Intel CAT and associate cores in the system to each partition such that a core is assigned to one or the other partition but not both. We refer to these partitions as *isolated partition* and *shared partition*. Processes belonging to containers designated as needing isolated execution will be pinned to the cores (using processor affinity) associated with the isolated cache partition and the rest will be pinned to cores associated with the shared cache partition. The maximum number of cache partitions available with Intel CAT is a hardware parameter and fixed for a given micro-architecture. The machine used for our testing allows for up to 4 distinct partitions, but newer machines have 16. But the configuration of size, number of active partitions, and core to partition assignment can occur in software and can be adjusted based on the demand for isolated execution and the needs of the expected workloads. Further, if there is no demand for shared execution then the host server could be partitioned into two (or more) isolated partitions.

As previously discussed, this hardware-assisted spatial partitioning protects the containers running in the isolated partition against cross-core prime+probe cache-side-channel attacks from containers running in the shared partition. However, cross-core cache-side-channel attacks across cache partitions are not entirely eliminated because Intel CAT, primarily designed to improve fairness of cache sharing and the performance of real-time workloads, allows cache hits across partition boundaries to maximize the benefits of shared memory (e.g., shared libraries). In particular, if the victim and the attacker processes have shared memory (e.g., because of layered file systems used in container frameworks), an attacker can flush cache lines associated with the shared memory of interest from his LLC partition, wait for a little while for the victim to execute, read the same shared memory and measure the time taken to see if there was a cache hit. Since the attacker previously flushed the cache line associated with the shared memory from his LLC partition, a cache hit indicates that the victim executing in a different core

---

[4]It can be extended to multiple isolated regions.

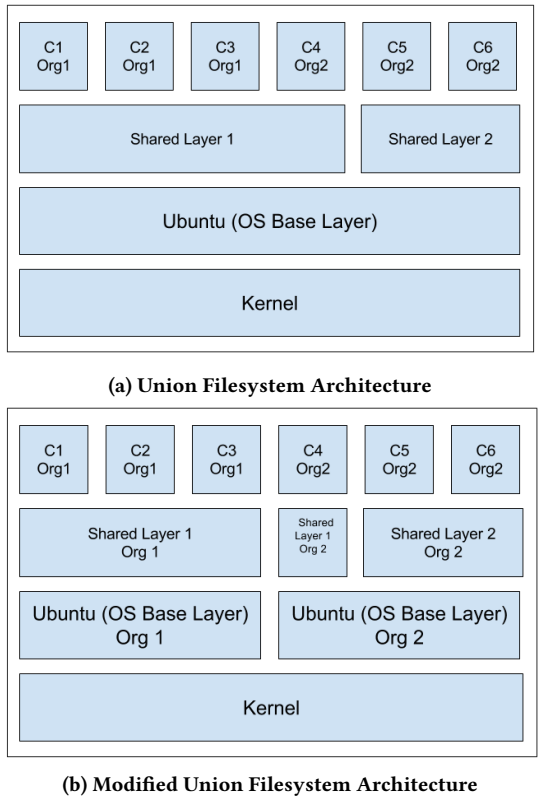

**(a) Union Filesystem Architecture**

**(b) Modified Union Filesystem Architecture**

**Figure 3: Defense against Shared-Memory Attacks**

and LLC partition has used or is using the library. While this limits the granularity of information an attacker can glean across partition boundaries, timing observations and hence the side-channel is not entirely eliminated. Further, cache-side-channel attacks from within an isolated partition continue to be viable. These will be addressed in the following subsections.

However, even the partial protection against cache-side channel attacks obtained through this spatial partitioning comes at the cost of reduced LLC cache size and the associated potential reduction in performance. Fortunately however, reduction in cache size has been shown to have relatively little impact on modern cloud workloads [12]. In particular, minimal performance sensitivity to LLC size has been reported above cache sizes of $4 − 6MB$ for modern scale-out and traditional server workloads (see Section 4.3 and Figure 4 in [12]) that are typical in cloud environments.

### 3.2 Selective Page Sharing

As previously discussed, hardware-assisted spatial partitioning does not eliminate cross-core flush-reload style cache-side channel attacks when the attacker and the victim have shared memory pages. Modern container deployments have one primary source of shared memory. Since Docker is one of the most popular choices for building container images and running them on Linux platforms, we limit our discussion to it, but these concepts are similar in other container frameworks. Docker uses storage drivers that are built on top of Union Filesystems (UFS) to present a filesystem to a process inside of a container from a stack of layers. Figure 3a shows how different layers are stacked and shared across different containers. Several different containers may use the same base components, thus a way was needed to reduce disk and memory usage of the common building blocks. Docker solves this problem by uniquely identifying each layer by its cryptographic hash and sharing the common ones between all containers built using a given layer ID.

Often there are multiple containers running the same image which causes them to share all the layers except for the upper most writable layer. For example, two Apache Tomcat servers running on the same Docker installation using the same image would share all the binaries including Java Virtual Machine (JVM), Apache Tomcat binary, GnuPG binary, and OpenSSL binary among others. Only the top most layer, containing writable elements such as Tomcat's log file, differ between containers.

To thwart flush-reload style attacks across cache partitions, we eliminate cross-domain page sharing through selective layer duplication. That is, for containers designated for isolated execution, our system allows sharing of layers only among containers belonging to the same tenant (or security domain) but not otherwise (see Figure 12b). This is a reasonable trade-off as it enables isolation between different tenants (or security domains) while limiting the increase in memory usage. In particular, the increase in memory usage will be a function of the number of tenants running on the server rather than the number of containers. We do not prevent traditional sharing of layers for containers running in the shared partition.

For VMs, kernel same-page merging (KSM) module in Linux used for memory deduplication is the main source of shared pages. However, KSM and memory de-duplication in general come with their own security risks(*e.g.*, [2, 3, 29, 33]). For instance it has been shown that KSM can be leveraged to break ASLR [2], can enable Rowhammer [20] attacks across VMs [29], and create a timing side-channel that can be used to detect the existence of software across VMs [33] much like the flush-reload style attack discussed previously. Given the serious security concerns surrounding the use of KSM we disable it in our framework.

Note, that selective page sharing combined with hardware-assisted spatial partitioning eliminate cross-core cache-side-channel attacks across partitions. Cache-side-channel attacks from within an isolated partition continue to be a threat and will be discussed next.

### 3.3 State Cleansing

Even with containers running in an isolated partition, an attacker allocated to the same isolated partition as the victim might be able to *(i)* observe the victim's LLC usage if scheduled to run on a different core than the victim but associated with the same partition, and *(ii)* even observe the victim's L1 and L2 usage if running on the same core as the victim (*e.g.*, [47]). In the latter case an attacker observes the cache usage of the victim by managing to frequently alternate execution with the victim process.

To thwart the latter kind of attacks we propose to cleanse the cache state when context switching between processes (containers) belonging to different tenants (or security domains). That is, if a process from one security domain (tenant), $SD_1$, runs on a core, then a processes belonging to another domain, $SD_2$, must either run on a core assigned to a separate partition or state-cleansing must be performed on the partition during the transition between process from $SD_1$ to process from $SD_2$. There currently exists no hardware instruction for per-partition cache invalidation. More details on how state cleansing is achieved are in Section 4. However, this does not prevent attacks from an attacker process who is running in parallel with the victim either on the same-core through simultaneous multi-threading or hyper-threading, or running on a different core but in the same partition.

A naïve solution would be to assign just one core to the isolated partition, disable hyper-threading and perform state-cleansing on every context-switch. The performance cost of such an approach is unattractive. A mitigation would be to create multiple isolated partitions with a single-core assigned to each. However, the number of cache partitions is finite (4 in our case), and such a an approach would further fragment the LLC for the shared partition. Further, many cloud workloads are multi-threaded.

## 3.4 Co-scheduling for Temporal Isolation

To address the aforementioned threat, we use a novel scheduling technique for temporal separation of security domains sharing a single cache partition. Co-scheduling container processes belonging to a given security domain across multiple processors amortizes the cost of state cleansing, but introduces additional complexity which we address below.

Scale-out workloads with many threads, those commonly deployed on cloud infrastructure, motivate this approach. As thread counts for a security domain increase, the number of threads able to run per domain at any given time remains high. This allows us to drive up utilization of cores assigned to a partition and only flush the partition when changing to the next domain. The complexities stem from needing to synchronize all isolated cores during domain changes, thus any implementation of co-scheduling has to guarantee an exclusion property. No task belonging to security domain $SD_X$ can run on an isolated processor while a task from another domain, $SD_Y$, is running in a processor associated with the same isolated partition. Additionally, before a task from $SD_X$ can run, a state cleansing event must occur. As shown in Figure 4, multiple cores can be utilized as once within a security domain, but then state-cleansing must be performed as every core assigned to a given partition context switches to another security domain. The next security domain cannot run on any isolated processor until this process is complete.

## 4 IMPLEMENTATION

Partitioning the LLC and associating cores with each partition does not need changes to the kernel or the operating system. It can be done by a system administrator as part of the machine configuration. Here we focus on the implementation of rest of the components.

## 4.1 Co-Scheduling

Co-scheduling *can* enforce isolation between security domains, but any implementation must be precise. By precise, we mean that any form of "loose" or "lazy" co-scheduling is unacceptable. For example,

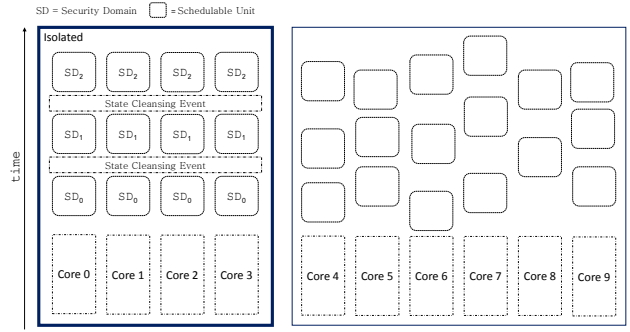

**Figure 4: Co-scheduling Overview - The isolated environment is on the left. It consist of an isolated cache partition along with the processors assigned to that partition. Co-scheduling is used to group tasks belonging to the same security domain and state cleasing events occur when changing domains. Regular tasks are on the right in a separate cache partition. Tasks on the right have no scheduling restrictions.**

consider a naïve implementation of co-scheduling as outlined in Figure 5.

**Table 1: Per Security Domain Thread Allocations**

| Security Domain | Thread Count |
|-----------------|--------------|
| ORG1            | 2            |
| ORG2            | 3            |

Figure 5 is a schedule instance of the configuration as described in Table 1. The example is a situation in which 2 cores are associated with an isolated partition and are running containers belonging to two security domains. These cores may be two physical cores or one physical core presented as two to the operating system (SMT). The defining characteristic in our example is the shared cache. For hypthreaded cores, this is the L1, L2, *and* LLC. In the case of two physical cores, the shared cache is only the LLC. The cores 1 and 2 in our example *are not* cores on two separate sockets on the same motherboard.

Consider a situation in which $Core_1$ initiates a domain transfer upon scheduling a thread from a conflicting domain, ORG2:THREAD1 in this example. Even if the scheduler invokes a flushing event, $f_1$, there remains a $\Delta t_3$ during which cross-core, cross-domain attacks could be carried out. This is seen again after $Core_1$ schedules ORG1:THREAD1 and ORG2:THREAD3 leading to durations $\Delta t_4$ and $\Delta t_5$ during which attacks remain feasible. While this situation is still better than the behavior of the default scheduler, it is not guaranteed to eliminate observable cache interference.

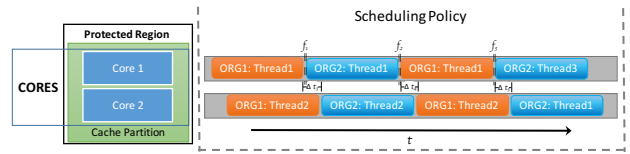

**Figure 5: Limitations of Naïve Co-Scheduling**

This example highlights the need for careful attention to details when implementing co-scheduling for improved security on modern operating systems. Effective elimination of side-channels dictates that cross-core synchronization be performed before state cleansing occurs and subsequent domain scheduling takes place.

To ensure that no two schedulable units belonging to different security domains run on an isolated cache partition simultaneously, we implement the core synchronization protocol shown in Figure 6. The protocol works by making the first core in an isolated partition a leader core. The leader core is responsible for initiating domain changes, synchronizing cores, and flushing the cache (state cleansing). All isolated cores only schedule tasks belonging to the `ACTIVE_SECURITY_DOMAIN`. Note that while only 2 cores are shown in Figure 6, the approach works with any number of cores. In Section 5 we evaluate the protocol with 4 cores assigned to an isolated partition.

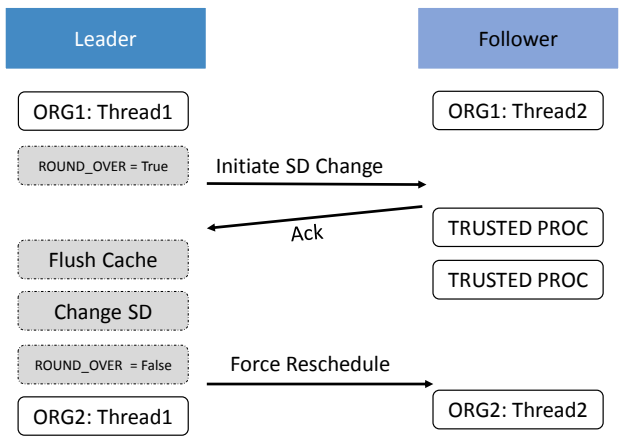

**Figure 6: Strict Co-Scheduling Protocol**

Isolated cores rely on two pieces of shared state to achieve strict synchronization. The leader core is the only core that can modify the state. The `ROUND_OVER` variable indicates to follower cores that a domain change is about to occur. A timer on the leader core initiates a domain change by modifying this variable and invoking the `__schedule` function on the leader core. The change domain event fires every `sysctl_sched_min_granularity`. After setting the `ROUND_OVER` variable to `true`, the leader core issues a reschedule command to follower cores and waits for them to send back an acknowledgment. The acknowledgment is performed within the `__schedule` function on follower cores. When the `ROUND_OVER` variable is set, partitioned cores can only run *trusted processes*. These are only kernel tasks, including: `ksoftirq`, `watchdog`, and the idle task.

After receiving an acknowledgment back from all follower cores, the leader then flushes the cache and updates the `ACTIVE_SECURITY_DOMAIN` to point to the next security domain. Our system uses a separate `task_group` in the Linux kernel for each security domain. All `task_groups` at the same level are linked together using a linked list. We use this list to implement a round robin style iteration through security domains on the leader core. Run-queue checking is performed to ensure a domain with runnable tasks is chosen.

Having chosen the next domain, the leader core sets `ROUND_-OVER` to `false` and again issues a reschedule command to follower cores. The `__schedule` function will eventually be invoked on the follower cores, but we use the reschedule command to reduce the idle time of follower cores. This protocol corrects the problem presented in Figure 5 resulting in "strict" co-scheduling as seen in Figure 7.

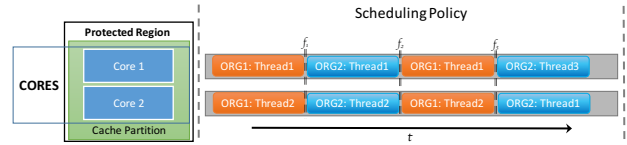

**Figure 7: Synchronized Co-Scheduling**

## 4.2 State Cleansing

The isolated cache partition must be cleaned or flushed before switching context to a different security domain. For a processor cache, state cleansing or flushing equates to invalidating the cache lines or evicting them, but no hardware mechanism exists to flush the cache lines assigned to a single CAT partition. The *WBINVD* instruction invalidates the entire shared cache disrupting processes in all partitions.

One way to implement state cleansing is for the user process to invoke the *CLFLUSH* instruction, which can evict cache lines corresponding to a linear virtual address and can be invoked from user space. This can be done by the application process before being switched out as was done in [50]. However, this requires changes to user applications which is not desirable. Another possibility is for the kernel to invoke *CLFLUSH* on the entire virtual address space. While this is guaranteed to work across processor generations, this approach is too costly. An optimization is to do it only on valid virtual addresses for the task being switched out as was done in [42]. However, this can still be a large range compared to the size of a cache partition (4 − 6MB).

Another approach is to create an *eviction set* – a set of addresses which when loaded are guaranteed to evict the entire cache partition. However, the memory address to cache line mapping is proprietary and subject to change across processor generations. Cache-side-channel attacks also have to contend with this challenge and have addressed it by reverse-engineering the memory address to cache line mapping for a given micro-architecture (*e.g.*, [23]). Apart from the one-time of cost of reverse engineering, the cost of this approach is equal to loading memory the size of cache partition. We adopt this approach.

We perform this state cleansing anytime the security domain is changed, as shown by the protocol in Figure 6 and the co-scheduling overview in Figure 4. To reduce performance impact in the case of other domains lacking runnable threads (due to blocking on I/O, *etc.*), flushing is only performed when `ACTIVE_SECURITY_DOMAIN` changes.

### 4.3 Selective Page Sharing

Docker uses Union File systems(UFS) to present a unified view of the several different layers. Of the several UFS that Docker supports like btrfs[9], overlayfs[10], AUFS[8], AUFS is one of the most mature one. In a UFS, multiple directories on the host are unified in a single directory called *union mount*, without replicating the actual contents of individual directories. Contents (files or directories) of all the directories become visible at *union mount*. Docker keeps single copy of each layer on the host filesystem and AUFS mounts all the layers to a single *union mount* point, which is then handed over to the container as its root file system. Each layer could be a part of multiple *union mounts* and thus can be shared across different containers.An important point to note here is that this behavior is specific to AUFS and Docker.

Our implementation modifies Docker (v1.14.0-dev, compiled from source code available on Github), specifically the AUFS storage driver, to transparently allow selective sharing of file system layers. Docker's command line client provides an option called `--cgroup-parent` which sets the parent cgroup of the container to the value provided. We modified the AUFS driver to have separate copies of each layer for each cgroup-parent. Each cgroup-parent maps to a separate security domain and thus no two containers belonging to different security domains have any common layers between them.

## 5 PERFORMANCE EVALUATION

### 5.1 Impact of Scheduler Changes

Feasibility is evaluated using a CPU bound workload to determine the impact on applications in the worst case scenario. Consider a batch workload such as Hadoop or a web serving workload. The case in which all threads have work and are not waiting for input is evaluated here.

The machine is configured as outlined in Section 2. We allocate 2 physical cores and 4 logical processors to an isolated cache region. The cores IDs are 4,20,5, and 21 due to the way Linux numbers logical processors. The cache region is 4MB (4 cache ways out of 20 available on the system). Each security domain is assigned 4 threads, and the number of domains is varied from 2 to 8. Each domain consists of 4 cpu bound tasks, 1 for each logical processor. Measurements are taken using `sar` and `pidstat` at an interval of once per second for 100 seconds. Figures 9-11 show where processor time is spent under different scheduler configurations. It is clear to see that the overheads for a single logical processor are a function of the system and not of the number of security domains assigned to a partition. . Follower cores in (b) and (c) can be seen idling during domain changes, but the overheads never exceed slightly above 10%, with the average case being slightly below 10% on follower processors. Flushing significantly increases the performance penalties as can been seen by the differences in (c) and (b) in each of the figures. The leader core spends the most time executing in system space due to its responsibility to change domains and synchronize cores, so this was to be expected. In the future, we will investigate mechanisms to reduce system time on the leader core and idle time on follower cores. Figure 8b shows the overhead normalized to security domain (each domain representing an organization with names testing[1-4] in this case) for the 4 security domain case. The reduction in time spent in userspace per domain is small.

### 5.2 Impact of Shared Memory Reduction

By enabling selective sharing of base layers in Docker, we expect an increase in the memory footprint of containers as there are multiple copies of certain pages that would otherwise be shared. To understand the memory growth vs. the number of security domains, we ran 2 experiments each with a web server (Apache Tomcat) and an in-memory database (Redis). We used `smem` to measure the proportional set size (PSS) per container as it represents realistic memory usage by only adding the fair share of the total shared memory. To make sure that the code is resident in memory before we take measurements, we sent 100 requests each to the Apache Tomcat servers and added 1000 random key-value pairs to each Redis server. We designed our experiments to be representative of real world deployments of micro-services where multiple containers each of a single type run in a distributed fashion.

In the first experiment we measured how the average memory usage of each container increased as we increased the number

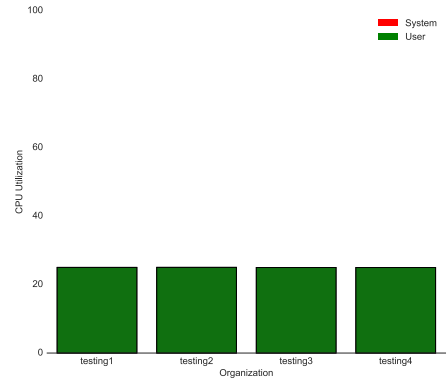

**(a) Baseline**

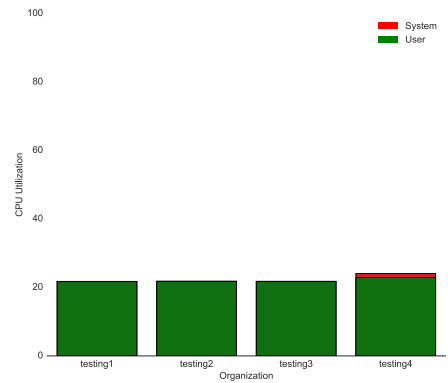

**(b) Co-Scheduling, Flushing**

**Figure 8: Overheads Per Domain**

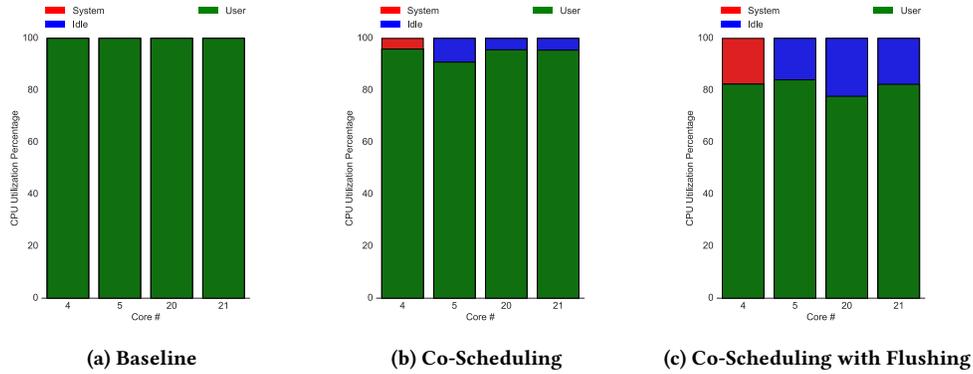

**(a) Baseline**      **(b) Co-Scheduling**      **(c) Co-Scheduling with Flushing**

**Figure 9: 2 Security Domains, 4 Logical Processors in Isolated Partition - Per Core Utilization**

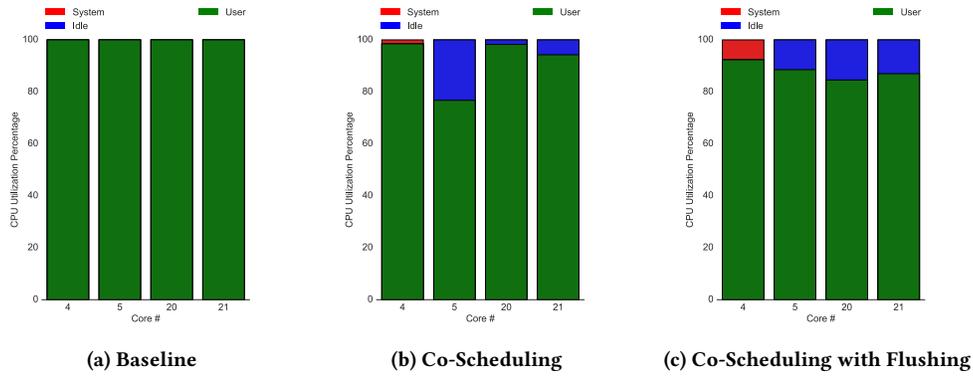

**(a) Baseline**      **(b) Co-Scheduling**      **(c) Co-Scheduling with Flushing**

**Figure 10: 4 Security Domains, 4 Logical Processors in Isolated Partition - Per Core Utilization**

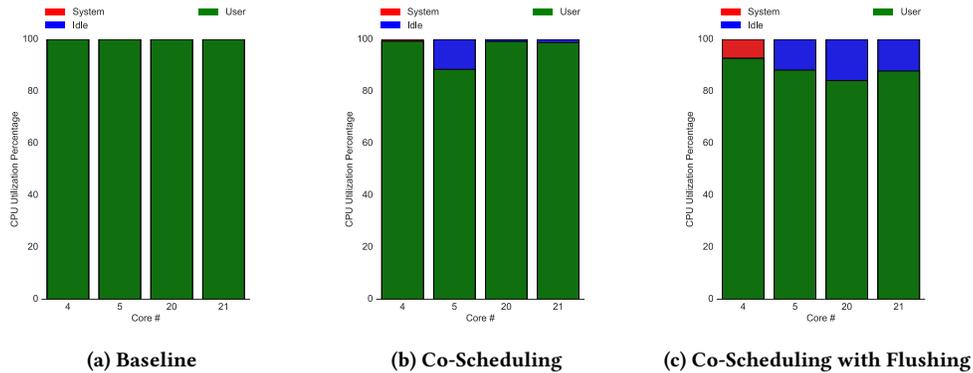

**(a) Baseline**      **(b) Co-Scheduling**      **(c) Co-Scheduling with Flushing**

**Figure 11: 8 Security Domains, 4 Logical Processors in Isolated Partition - Per Core Utilization**

of security domains from 1 to 4. Each security domain owns 5 containers. It was compared against the same number of containers running without modifications on Docker (*i.e.,* all the containers share base layers). We measure the average memory usage for across 50 runs. Figure 12a and 12b show that memory usage for Redis increased only about 0.45 MB per container and for Apache Tomcat only about 1.71 MB.

In the second experiment we ran 20 containers equally split between 1, 2 and 4 security domains and observed the increase of memory usage per container. Figure 12c and 12d show that memory usage for Redis increased only about 1.61 MB per container and for Apache tomcat 3.26 MB.

The results support our claim that the additional cost of selective sharing of memory is negligible. The increase that we see arises from the duplicate copies of memory pages, one per security domain, which were shared among all the containers within that domain.

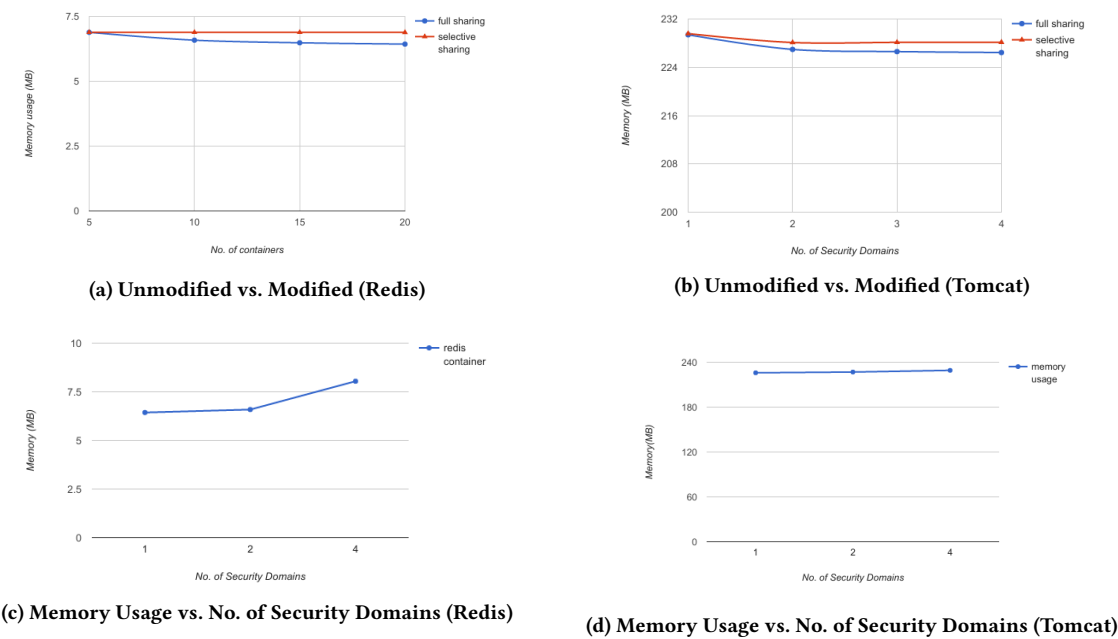

(a) Unmodified vs. Modified (Redis)

(b) Unmodified vs. Modified (Tomcat)

(c) Memory Usage vs. No. of Security Domains (Redis)

(d) Memory Usage vs. No. of Security Domains (Tomcat)

Figure 12: Memory Overheads

# 6 RELATED WORKS

## 6.1 Cache Side-Channels

Cache side-channel attacks take advantage of the shared nature of processor resources, in particular the processor's Level-1, Level-2, and Last-Level caches. Prime+Probe attacks were first explored across Virtual Machines (VMs) by Osvik *et al.* [27] and shown to be practical in cross-core attacks via the LLC by Liu *et al.* [23]. Modern clouds are driving up machine utilization by offering container based platforms. This increases revenue [36] while providing more performance to tenants [11, 41]. Cache side-channel attacks are already emerging on container based infrastructure [48]. Flush+Reload attacks [18] are a real threat to cloud computing security and have been successfully deployed on public infrastructure [48]. All of these attacks fall under the broader class of *access based* side-channels in which an attacker can tell whether or not a given resource has been accessed by a victim in a given time period.

## 6.2 Existing Solutions

Existing approaches achieve single core isolation by disabling hyperthreading. StealthMem is able to enable hyperthreading, but the authors do not sufficiently address cross thread scheduling issues [19] and it requires application developers to make code changes. Disabling hyperthreading significantly reduces the throughput of not only the "secure" workload, but the entire machine. Disabling hypthreading is untenable as the economic model behind cloud computing dictates high per machine utilization [36]. Cutting whole-machine utilization by even 20% (the impact of hyperthreading in 2005 [4]) is too high for cloud computing. We argue that for some tenants, a 20% overhead may be a reasonable trade-off for increased security, there is little value in forcing all tenants pay that performance penalty. CATalyst [22] follows a similar defense model to Stealthmem but uses Intel's CAT technology to assign virtual pages to sensitive variables instead of software-based page coloring.

CACHEBAR [51] defends against Flush+Reload attacks by duplicating memory pages on access from separate processes, a scheme it calls Copy-On-Read. Since Linux' KSM does de-duplication of pages at regular intervals, it also modifies the behavior of KSM to achieve Copy-On-Read. To defend against Prime & Probe attack CACHEBAR modifies the memory allocation based on which cache lines the memory regions maps to so that the attacking process loses visibility into the victim process. It provides only probabilistic guarantees for defense against a Prime & Probe attack.

Some solutions are probabilistic [25, 37, 51] or do not protect applications when SMT is enabled [51]. Others require developers to re-write applications [19, 22, 31] or rely on costly hardware changes [39, 40]. Other hardware approaches violate x86 semantics by modifying the resolution, accuracy or availability of timing instructions [21, 24, 38] . Finally, cache coloring [28, 32, 45] approaches have impractically high overheads. Solutions like Nomad[25], while probabilistic, complement our approach. Nomad works in the cloud scheduler to reduce the co-residency of different security domains. Our solution could be used in conjunction with Nomad to provide hard isolation when co-residency restrictions are not possible.

# 7 CONCLUSIONS

We have presented a hardware/software mechanism that eliminates cache based side-channels for schedulable units belonging to separate security domains. Unlike many existing solutions, our solution allows SMT to remain enabled and does not require application level changes. A user simply notifies the provider that a given workload should be run in isolation. Our solution eliminates an attackers ability to use the cache as a noisy communication channels and does not rely on probabilistic methods to decrease the granularity of information available on the channel. We implemented our system on top of the Linux scheduler and presented an evaluation of the system under a CPU bound workload. Our system has a worst case reduction in utilization in the case of 2 security domains of 9.8% and only 2.97% and 1.68% decrease in utilization for the 4 and 8 security domain configurations respectively (with 4 logical processors assigned to an isolated partition).